\documentclass{aa}
\usepackage{graphicx}
\usepackage{txfonts}
\begin{document}
\title{Coronae above accretion disks around black holes: The effect of Compton cooling}


\author{E. Meyer-Hofmeister\inst{1}, B.F. Liu\inst{2} and F. Meyer\inst{1}}
\offprints{Emmi Meyer-Hofmeister; emm@mpa-garching.mpg.de}
\institute
    {Max-Planck-Institut f\"ur Astrophysik, Karl-
     Schwarzschildstr.~1, D-85740 Garching, Germany
\and National Astronomical Observatories, Chinese Academy of Sciences, 
20A Datun Road, Chaoyang District, Beijing 100012, China} 
     \mail{emm@mpa-garching.mpg.de}

\date{Received: / Accepted:}
\abstract
   {The geometry of the accretion flow around stellar mass and
     supermassive black holes depends on the accretion rate. Broad
     iron emission lines originating from the irradiation of cool matter
     can indicate that there is an inner disk below a hot coronal flow.}
    {These emission lines have been detected in X-ray binaries. 
     Observations with 
     the {\it{Chandra X-ray Observatory, XMM Newton}}
     {\rm {and}} {\it{Suzaku}} {\rm{ }} have confirmed the 
     presence of these emission lines also in a large fraction of Seyfert-1
     active galactic nuclei (AGN). We
     investigate the accretion flow geometry for which broad iron emission
     lines can arise in hard and soft spectral state.}
    {We study an ADAF-type coronal flow,  where the ions are viscously
      heated and electrons receive their heat only by collisions from
      the ions and are Compton cooled by photons from an underlying
      cool disk.}
    {For a strong mass flow in the disk and the resulting strong Compton
      cooling only a very weak coronal flow is possible. This
      limitation allows the formation of ADAF-type coronae above weak
      inner disks in the hard
      state, but almost rules them out in the soft state.}
      {The observed hard X-ray luminosity in the soft state, of up to 10\% or
      more of the total flux, indicates that there is a heating
      process that directly accelerates the electrons. This might point to
      the action of magnetic flares of disk magnetic fields reaching
      into the corona. Such flares have also been proposed by
      observations of the spectra of X-ray black hole binaries without
      a thermal cut-off around 200 keV.}




\titlerunning {Accretion disks coronae: 
Compton cooling}

\maketitle
%

\section{Introduction}
With the advent of {\it{XMM-Newton}}
observations it has become obvious that broad iron emission lines are
common in Seyfert galaxies (Nandra et al. 2007). The number of
sources with detections of strong disk lines
is growing with the increasing number of new observations by
{\it{XMM-Newton}}, {\it{Chandra}} and {\it{Suzaku}}. These Fe K-shell
emission lines are the strongest X-ray emission lines in most AGN and
X-ray binaries (see the review by Miller 2007). Averaging AGN X-ray
spectra from deep {\it{Chandra}} fields Falocco et al. (2011)
significantly detected the broad component of the Fe line in the
X-ray spectra of low luminosity AGN at low redshift.

Since iron emission lines can be understood as the reaction of an
accretion disk to irradiation by an external source of hard X-rays, the
obvious question is under which conditions the accretion geometry
contains an inner disk and above it a hot coronal
flow/ADAF. An inner disk is necessary to explain the relativistic lines. As 
commonly accepted, the mass accretion rate determines
whether an optically thick and geometrically thin disk reaches inward to
the innermost stable circular orbit (ISCO) or the disk is truncated
and replaced by an advection-dominated optically thin hot flow in the 
inner region. A detailed definition of these states
is given by Remillard and McClintock (2006). For very
high accretion rates, a third, so-called steep-power law state can
exist, which is observed during some outbursts of X-ray binaries. Though in
principle the accretion flow geometry
is the same for X-ray binaries and AGN (Narayan et al. 1998), it is easier
to distinguish between different spectral states for
galactic sources than for AGN. Walton et al. (2012) analyzed high
quality spectra of the stellar-mass black hole binary XTE J1650-500 and
the active galaxy MCG-6-30-15 and argued that the similarity of broad
iron line features shows that the excess in both sources is of the same nature.

During the hard/low state, the power-law flux from the ADAF in the
inner region is the main source of radiation.
If the disk is truncated, no relativistic iron emission lines should be 
produced. However, from the earliest studies a broad iron emission line
has been detected
for Cygnus X-1 in low state ( Barr et al. 1985). Now {\it{Chandra}} and
{\it{XMM-Newton}} observations have firmly established the existence of 
a relativistic Fe K emission line in spectra of several X-ray
binaries. Reis et al. (2010) analyzed the X-ray spectra of
black hole binaries in the canonical
low/hard state and found that the accretion disk reaches inward to the
ISCO. These detections were made mostly during outburst decline and
can theoretically be understood in terms of re-condensation of gas into 
an inner disk (Meyer et al. 2007) in the hard spectral state, as a
transient phenomenon following the state transition. This accretion
flow geometry, an outer truncated disk, and inside only a weak inner
disk and above it the main flow via a corona/ADAF, is the situation
where broad iron emission lines are expected. There may be a gap
between the outer standard disk and such a weak inner disk.

During the high/soft (thermal) state, the flux is dominated by the
radiation from the inner disk. The observations of X-ray binaries
have identified an additional hard power-law flux, sometimes with a
non-thermal tail, the power-law flux that is evidence of the co-existence of
disk and hot flow. In this case then, one has to keep in mind that heat
conduction and the possible exchange of mass and angular momentum between
disk and hot flow can be significant. Investigations of the spectra of X-ray binaries, especially Cyg
X-1 (Ibragimov et al. 2005), led to the conclusion that magnetic
flares are needed to explain the non-thermal radiation.Hints to
magnetic fields also come from accretion disk winds, which are
ubiquitous in black hole binaries in the soft state, as pointed out by Ponti et
al. (2012). Miller et al. (2008) modeled spectra observed for GRO
J1655-4 and concluded that these disk winds  are probably related to
magnetic fields.

The interpretation of spectra is less clear for AGN. Despite
the large number of observations, it is difficult to classify the source
as in hard or soft state. Vasudevan \& Fabian (2009) presented 
spectral  energy distributions for a sample of AGN, based on
contemporaneous optical, ultraviolet, and X-ray observations, and
determined the ratio of bolometric to Eddington luminosity, using
the black hole mass measurements of Peterson et al. (2004). Tanaka et al. (1995)
first observed an asymmetric disk line profile in the Seyfert-1 AGN 
MCG-6-30-15, which remains a very important source for studies of
relativistic disk lines. If a
relativistic line is detected, one might conclude that an inner disk
exists, which is either (1) in the soft state, a standard disk inward
to the ISCO with
irradiation from a coronal layer or (2) in the hard state, only a very weak 
inner disk below the standard ADAF. The latter case seems to occur in 
X-ray binaries as a transient phenomenon of the intermediate spectral
state. One might expect to find  the same accretion flow geometry for
AGN. Meyer-Hofmeister \&
Meyer (2011) investigated the situation 
using Seyfert galaxies for which emission lines were clearly present
(Nandra et al. 2007, Miller 2007), but found that some of these
sources were apparently in the soft rather than the hard state. 

That broad iron emission lines are common in Seyfert AGN 
suggests that they are not a transient phenomenon and that the accretion
geometry then corresponds to a soft state disk with a hot coronal 
layer/ADAF above, at least
for some of the observed sources. In this paper we investigate the structure of
the hot coronal flow above a standard disk under the
influence of Compton cooling. In Sect.2, we give a short summary of
broad iron emission lines detected in X-ray binaries and AGN. Sect.3
concerns the physics of interaction of disk and corona around
supermassive and stellar-size black holes. We determine
the limitation of the coronal flow caused by
Compton cooling (Sect.4). The restriction in the case of a strong disk
mass flow is so low that the observed hard power-law flux in a number
of sources requires an additional heating process, e.g. 
one produced by magnetic flares. This is consistent with
the conclusions derived by Ibragimov et al. (2005) from fitting  X-ray binary spectra with a
non-thermal tail and no cut-off at energies between 100 keV and 200
keV (Sect.5). Whether the situation is similar for
AGN is difficult to judge since for these sources, a non-thermal power-law
flux does not seem to be indicated, but cannot be not generally
excluded. In Sect.6,
we refer to observations for X-ray binaries that detect evidence of non-thermal
Comptonization. A discussion of questions related to our results
follows in Sect. 6.

\section{Detection of broad iron emission lines in Seyfert AGN and
  X-ray binaries}

Broad iron emission lines provide information on the
accretion process in the innermost region around the black hole.
If the line profile is asymmetric, the extent of the red
compared to the blue wing allows us to derive measures of the black
hole spin. Miller (2007) presented a review of relativistic X-ray
lines from both the inner accretion disks of Seyfert AGN and disks around 
stellar-\ size black holes. The recent {\it{XMM-Newton}} and 
{\it{Suzaku}} surveys were discussed, in which broad Fe lines were found to be
common. Nandra et al. (2007) found broad iron emission lines for 73\%
of the sources in his sample of 30 Seyfert galaxies observed with
{\it{XMM-Newton}}. Bhayani
\& Nandra (2011) achieved an improved fit of these data, finding evidence of
broad iron lines in 11 sources for which originally the lines could
not be detected. Reeves et al. (2006) report the detection of relativistic
disk lines and reflection in six of seven Seyfert AGN in a
{\it{Suzaku}} survey. These observations also show
that the disk reflection spectrum
follows the same variability pattern as the iron line (see discussion
in Miller 2007). The much cooler accretion disks in AGN allow us to
easily separate disk radiation from the iron emission line in the
spectrum. In X-ray binaries, the radiation from the disk and the
corona, which in both cases both lies in the range of commonly observed X-rays and iron
emission lines would be more difficult to be recognized.

A priori it is unclear what the detection of so
many iron emission lines in Seyfert galaxies means with respect to the accretion
flow geometry. Brighter sources with high mass accretion rates in
the high/soft spectral state may be overrepresented in the samples. 
However, Nandra et al. (2007) found a
mean spectral index of $\Gamma$=1.86 for their sample, which according
to the definition of Remillard \& McClintock (2006) would correspond 
to a low/hard state. 

In X-ray binaries, the spectral state can usually be determined. In
these systems, the broad iron emission lines 
are primarily found in the low/hard spectral state, either during a rise to an
outburst or, more often, an outburst decline, at times when the
accretion rate has not yet decreased to very low values below about 0.001
of the Eddington rate. These lines were detected in connection with very
weak inner disk components. Hence the iron emission lines appear to be
a transient phenomenon, that has been found for quite a number of well-known X-ray binaries 
(for compilations of the detections, see Miller 2007, Meyer-Hofmeister
et al. 2009, Reis 2010). These iron lines seem to require cool
material below a hot flow close to the central black hole. This has
sometimes been taken as a contradiction to the ``truncated disk
picture'' in the hard state. However, these weak inner disks could
well form by the re-condensation of gas from the hot flow
(Liu et al. 2006, Meyer et al. 2007, Taam et al. 2008).
For some Seyfert AGN with observed broad iron emission lines, the
accretion flow geometry might be the same. On the other hand, 
``best candidates'', for the presence of broad iron emission 
lines in Nandra et al. (2007) and Miller (2007)
seem to have an Eddington scaled accretion rate above 0.1, hence
are probably in the soft spectral state (Meyer-Hofmeister \& Meyer 2011).

\section{Advection-dominated accretion flows}

The origin of hard X-rays observed in accretion flows onto compact objects is 
commonly attributed to the inverse Compton scattering of cool photons from
outside on hot electrons in the accreting plasma. The high electron 
temperature results naturally from the effective heating and rather poor 
radiative cooling of the
tenuous plasma. Major heating processes here are the collisional heating
by hot ions generated by the ``frictional'' release of gravitational
energy in the accreting flow, the dissipation of magnetic energy in the
corona of an underlying accretion disk, and conceivably the strong
irradiation by an external source of high energy photons.

An advection-dominated optically thin hot flow, which was discussed by 
Ichimaru (1977) and both introduced and investigated in detail by 
Narayan and collaborators (see
Narayan et al. 1998 for references), is presumably present in many X-ray
sources, stellar-mass objects and AGN. The hard spectrum mainly
results from the hot plasma in an inner region around the black hole. 
The appearance of magnetic
fields is nearly unavoidable in accretion flows of conductive plasma
and in particular they are thought to provide the frictional heating in
accretion disks by the magneto-rotational instability. These fields may
well reach out to an overlying corona and there produce magnetic
heating e.g. by flares. Irradiation from a central source can heat the corona 
above the disk, as in the case of Her X-1 (Schandl \& Meyer 1994). 

\subsection{Interaction between corona and disk}

An inner hot flow  is usually surrounded by a standard Shakura-Sunyaev
type accretion disk. The change from disk accretion to a hot flow
 can arise from thermal conduction from a hot corona, that
evaporates the disk underneath in a way that depends
on the accretion rate (Meyer \& Meyer-Hofmeister 1994, Liu
et al. 2002). In a typical low/hard state with accretion rates below a 
few percent of the Eddington accretion rate ($\dot M_{\rm{Edd}}$), the
inner disk is evaporated and recedes outward to distances of several hundred
Schwarzschild radii. 

The opposite process, the condensation of gas from the hot flow onto a
cool disk below, is also possible (Liu et al. 2006, Meyer et
al. 2007) and can help to sustain a very weak inner disk after an
outburst. In such a hard-intermediate state, accretion rates are around
0.001 to 0.01 $\dot M_{\rm{Edd}}$. An inner weak disk can explain
the broad Fe K emission lines in the spectra of X-ray binaries in
the hard spectral state. The modeling is supported by an analysis of 
observations for  GX 339-4 (Liu et al. 2007, Taam et al. 2008).
A hot corona above a very weak cool disk in a hard/low spectral 
state differs
from a pure ADAF: The existence of a cool disk underneath the
corona by vertical thermal conduction sets up an electron
temperature profile connecting the hot corona and cool disk, and the
electron temperature near the base of the corona finally becomes so low
that Coulomb coupling between the ions and electrons becomes effective 
and ion and electron temperatures become equal.

In the high/soft spectral state, if the
underlying disk is strong, another case of an interaction between disk
and corona is present. The inverse Compton scattering of cool disk photons
by hot electrons leads to an efficient cooling of the electrons. We
study under which circumstances this process can completely cool the corona.

In the following, we distinguish between a standard ADAF 
present in the inner region in the hard spectral state and
an ``ADAF-type corona'' as a coronal flow above a weak or strong disk.
Common to both cases is that only the ions are thought to be directly
heated by the release of gravitational energy in the accreting gas, and the
electrons only receive their heat by collisional coupling to the ions.

\subsection{The two-temperature ADAF-type corona in the inner region}

A two-temperature flow is present if electrons are cooled
preferentially. In general, the collisions between ions and electrons
reduce a temperature difference. At large distances, larger than a few hundred
Schwarzschild radii, ions and electrons are thermally coupled. 

We consider here the changes that occur to the structure of the
coronal flow when a
disk exists below the hot flow. If there is no additional heat input
from a disk underneath, then such an ``ADAF-type corona'' will be similar to a
standard ADAF where the electrons receive their heat only from collisions 
with the hot ions, which in turn have been heated by the release of
potential energy. We use the analytical relations of Narayan \& Yi (1995b). 
Total pressure $p$, electron density $n_e$, viscous dissipation of
energy $q^+$, and isothermal sound speed $c_s$ 
depend on the black hole mass $m$, the accretion rate $\dot m_c$ in the ADAF,
 the distance $r$ from the black hole, and the viscosity
parameter $\alpha$ as

\begin{eqnarray}\label{ADAF}
p&=&1.14\times 10^{16}\alpha^{-1}m^{-1}c_3^{-1/2}\dot m_c r^{-5/2}
\, \rm{g\ cm^{-1} s^{-2}} \nonumber \\
n_e & =&1.33\times10^{19}\alpha^{-1}m^{-1}c_3^{-3/2} \dot m_c r^{-3/2} \,  \rm{cm^{-3}} \\
q^+ &=& 1.84\times10^{21}\epsilon'c_3^{1/2} m^{-2}\dot m_c r^{-4}\rm{ergs\ cm^{-3} s^{-1}}
\nonumber \\
c_{s}^2&=& 4.50\times 10^{20} c_3 r^{-1} \rm{cm^2 s^{-2}}, \nonumber 
\end{eqnarray}
where $m$, $\dot m_c$, and $r$ are in units of solar mass, Eddington
rate ($\dot M_{\rm Edd} =1.39\times 10^{18}m$ g/s), and
Schwarzschild radius, respectively. We denote by $\dot m_c$ the ADAF
accretion rate because we later introduce $\dot m_d$, the accretion
rate in a disk beneath the ADAF. The coefficients depend on the
ratio of specific heats of the magnetized plasma $\gamma$

\begin{eqnarray}
c_3=\frac{10+4{\epsilon}'}{9\alpha^2}\left[ \left(
  1+\frac{18\alpha^2}{(5+2\epsilon')^2} \right)^{1/2} -1
  \right]\approx {2\over 5+2\epsilon'} , \hspace {0.5cm}
\end{eqnarray}
with
\begin{equation}
\epsilon'=\frac{1}{f} \left( \frac{5/3-\gamma}{\gamma -1}\right).
\end{equation}
The parameter 
$f=(q^+-q^-)/q^+$, which is the ratio of the advected energy
to heat generated, measures the degree to which the flow is advection-
dominated, where $q^-$ is the energy lost. This means that in a
standard ADAF $f$ is close to 1.

The expressions depend on the
ratio of specific heats of the magnetized plasma
 $\gamma=(8-3\beta)/(6-3\beta)$ (Esin 1997), where $\beta$ is the
ratio of gas pressure to total pressure. We adopt the value $\beta$=0.8,
resulting from shearing box simulations in a collisionless
plasma (Sharma et al.2006).

As long as the ions are not significantly cooled by the transfer of
energy to the electrons they take their energy inward with the flow.
This determines the virial-like ADAF temperature of the ions. However, if the
ions lose a certain fraction of their energy to the 
electrons, their temperature becomes accordingly lower. For a
sufficiently low electron temperature, electrons and ions
become thermally coupled.
The rate of collisional energy transfer from ions to
electrons given by Stepney (1983), is approximated (Liu
et al. 2002) as

\begin{eqnarray}\label {qie}
q_{ie}&=&A \, n_e n_i \frac{T_i-T_e} {T_e^{3/2}}  \\ 
A&=&1.639\times 10^{-17}  \rm{g\ cm^{5} s^{-3} deg^{1/2}}  \nonumber
\end{eqnarray}
where $ T_e$, $T_i$ are the electron and ion temperatures, $n_e$ and $n_i$ 
are the electron and ion number density. 

This energy transfer increases with decreasing electron temperature. When the 
collisional heat transfer from ions to electrons is able to take away nearly
all the heat that the ions have gained, i.e. $q_{ie}=(1-f)q^+$, $f$ small,  coupling occurs. 

We use the
analytical relations for the case $f=0.1$ (as shown later an
appropriate approximation). The electron 
temperature at which coupling occurs can then be roughly estimated to be

 \begin{equation}\label{T_cpl}
T_{\rm {cpl}}=  3.30\times 10^9 \alpha^{-4/3}\dot m_c^{2/3} K,
\end{equation}
which is independent of distance $r$.

We note that a low electron temperature can also result from electron
cooling by thermal conduction to the underlying
disk. These coupling processes were discussed in the context of the re-condensation
of gas onto a weak inner disk (Meyer et al. 2007). 

\subsection {The relation between $T_i$ and $T_e$}

The ion temperature is the result of viscous heating and collisional
cooling by electrons. This collisional cooling depends only on the
ADAF parameters and the electron temperature and is independent of the
nature of the electron cooling processes. The relation between ion and
electron temperature describes the possible combinations of the two 
temperatures in the case of any cooling process leading to a temperature decrease.

The gas pressure of the accreting gas is given by 
\begin{equation}
\beta p = \rho\Re (\frac{T_i}{\mu_i} +\frac{T_e}{\mu_e}) 
\end{equation} 
with $\mu_i$=1.23 and $\mu_e$=1.14 for a
hydrogen mass fraction of 0.75.  From Eq.(\ref{ADAF}) we derive a
relation between $T_i$ and $T_e$
\begin{equation}\label{e:Ti-Te}
T_i+1.08T_e=6.66\times 10^{12}\beta c_3 r^{-1} \rm{K}.
\end{equation}
In the  above equation
$c_3$ contains the parameter $f$, which is determined by  $f=1-q_{ie}/q^+$. Thus, $c_3$ is an implicit  function of  $T_e$, $T_i$, $n_e$, and
$n_i$.  Given $\alpha$, $\beta$,$\dot m_c$, and distance $r$,
the relation between $T_i$
and $T_e$ is the solution of the implicit equation (\ref{e:Ti-Te}). 

\begin{figure}
   \centering
   \includegraphics[width=6cm,height=5cm]{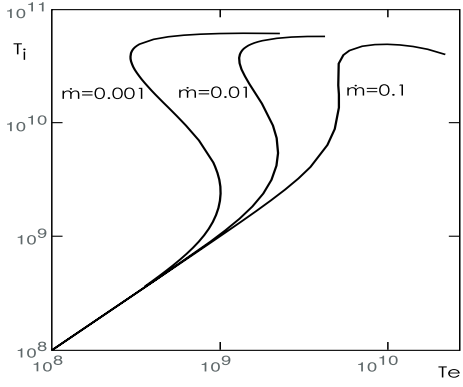}
      \caption{ 
The change of ion temperature with electron
        temperature for given rates of coronal mass flow $\dot m_c$
        at distance $r=30$, 
        $\alpha$=0.2 and $\beta$=0.8} 
   \label{f:logTi-logTe1}

   \end{figure}

 \begin{figure}
   \centering
   \includegraphics[width=6cm,height=5cm]{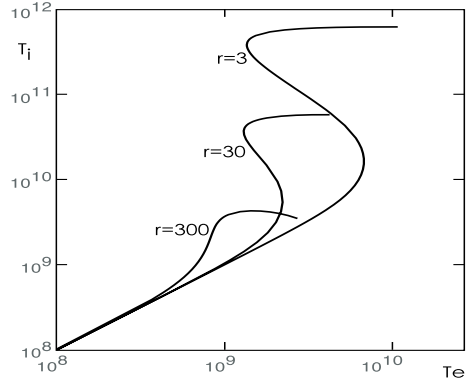}
       \caption{Relation between ion and electron temperature, 
               for different distances from the black hole, and $\dot
               m_c$=0.01, $\alpha$=0.2.}
         \label{f:logTi-logTe2}
   \end{figure}

Fig.\ref{f:logTi-logTe1} shows the relation for various mass flow
rates $\dot m_c$ in an ADAF at distance $r$=30. Three branches are
displayed. A first branch in the uppermost
part of the figure describes the standard ADAF: the ion temperature is
high, ions and electrons are thermally decoupled, and the ions obtain the 
virial-like temperature of an ADAF. 
If the electron temperature decreases owing to a cooling process, 
collisional coupling increases, more heat is drained from the ions,
the ion temperature starts to decrease, and the electron
temperature  finally reaches a minimum. A second branch of possible
combinations of values $T_i$ and $T_e$ then
appears: the ion temperature decreases further, and the
electron temperature rises again. This is due to the rate of collisional
energy transfer from ions to electrons for these $T_i$, $T_e$ values
and the corresponding density values (Eq.\ref{qie}). This second branch
is an unstable structure, corresponding to the solution found by
Shapiro, Lightman and Eardley 1976), the instability depending on the cooling
process (Piran 1978). At sufficiently low ion temperature
a third branch then appears where the ion and electron
temperature become equal. These three branches correspond to
the possible configurations for the different cooling processes of electrons
and ions, should not be considered as consecutive structures. In the
following section, we discuss the effect of especially Compton cooling.

 The advected energy fraction $f$ decreases from 1 
at the upper right end of the $T_i$ - $T_e$ curve towards zero at the
one-temperature end, with $f$ about 0.1 at the upper bend of the curve. In the
limit of very small $f$, the corona is no longer spherical and the
density of the coronal gas increases strongly towards the equatorial
plain (Narayan \& Yi 1995a).

For the Compton cooling process, we discuss here only an optically thin
flow. The scattering optical depth of an ADAF 
similarity solution (Narayan \& Yi 1995b) is

\begin{equation}
\tau_{\rm{es}} = 8.27 \alpha^{-1}c_3^{-1} \dot m_c r^{-1/2},
\end{equation}
and becomes $\ge$ 1 for low values of $f$.

 We note that for
larger optical depth the cooling efficiency of the disk photons is 
higher and the derived limits on the allowed coronal flow
are even lower.
{\rm{ 

Fig.\ref{f:logTi-logTe2} shows the change in the relation with
distance.  $r$=30 is a typical distance for an inner region, a finding
that is important
for the radiation, which determines the spectrum. In the innermost
area, $r$=3 the temperatures might be affected by processes that are not
considered here. $r$=300 is already a distance where the ion
temperature is not very high compared to the electron temperature, 
indicating that there is a transition to the farther outward located one-temperature ADAF.

The relation between ion and electron temperatures in
Figs.\ref{f:logTi-logTe1} and \ref{f:logTi-logTe2} for small
distances shows an interesting hysteresis. When cooling increases,
e.g. owing to an increase in the accretion rate in the disk underneath,
the ion temperature can abruptly drop to the value $T_i=T_e$.
On the other hand, starting from a 
one-temperature flow, allowing temperatures to increase, the plasma 
remains at one temperature until it reaches a 
critical value where ions become uncoupled from the electrons,
cannot be cooled efficiently, and assume the virial-like
temperature of the two-temperature ADAF. The two transition
temperatures are different. The reason for the difference is that 
as long as the ions are coupled with electrons their temperature
remains low, the density is large, and the coupling remains strong.

\section{Compton cooling: A weak ADAF-type corona above a strong disk in
  soft spectral state}

  \begin{figure}
   \centering
   \includegraphics[width=6cm,height=5cm]{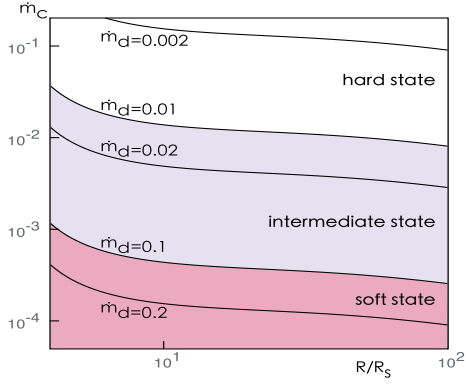}
      \caption   
{The effect of Compton cooling: Upper limit to the mass
        flow rate $\dot m_{\rm{c}}$ in the ADAF-corona for different 
        mass flow rates $\dot m_d$ in the disk and distances, 
        $\alpha$=0.2 
        (approximation for f=0.1). Rates $\dot m_d$=0.1-0.2 correspond to a
        typical soft state, and rates $\leq$0.01 to an intermediate state.}
   \label{f:mdot_c}
   \end{figure}

In a hot
advection-dominated coronal flow above an underlying disk conduction
cooling and Compton cooling  are in general the most important cooling
processes. When conduction
cooling is not important the electron temperature is determined by the
inverse Compton scattering by photons from the disk. The
Compton cooling rate is

\begin{equation}\label{qcmp}
q_{\rm {cmp}}=\frac{4kT_e-h\bar{\nu}} {m_e c^2} n_e\sigma_T\sum_i{u_ic},
\end{equation}
with $\sigma_T$ the Thomson scattering cross section, $\sigma$ the
Stefan
Boltzman constant, $h\bar{\nu}$ the mean photon energy, and
$u_i$ the energy density of the cooling/heating photons. 
There are three sources of Compton cooling: photons from the disk below,
photons from the innermost disk region and hard radiation from the
hot flow. In the first two cases the term  $h\bar{\nu}$ in
general can be neglected compared to the thermal electron energy.  The
contribution from the disk underneath is

\begin{eqnarray}
u_1c&=&2F_{d}, \\ \nonumber
F_{d}&=&\sigma T_{\rm{eff}}^4 ={3GM\dot M_{\rm d}\over 8\pi R^3}\left(1-\sqrt{3/r}\right)
\end{eqnarray}

where $u_i$ is the energy density of the photon field, $M$ the black hole mass, 
$\dot M_{\rm d}$ the disk accretion rate, $R$ the radius, and $R_S$
the Schwarzschild radius.

For the contribution of the inner disk region, we assume that it is diluted 
in proportion to $1/R^2$ at distances larger than about
$5R_S$. The inclination angle $\delta$ under which the assumed flat inner disk
appears at coronal height $z$ is given by $\tan \delta=R/z$. For a density
distribution $\rho \propto exp(-z^2/H^2)$ with scale height
$H=\sqrt{c_s^2/\Omega^2}$, for a rotational velocity $\Omega$, we
obtain a mean height $\overline{z}=H/\sqrt{\pi}$. With this mean height, the
corresponding projection of the disk area with respect to the line of
sight yields

\begin {equation}
u_2c={L_d\over {4\pi R^2}}{1\over {\sqrt{1+\left(\frac{R\Omega}{c_s}\pi\right)^2}}}.
\end{equation}
We take the central luminosity as $L_d=0.1\dot M_dc^2$. This leads to the sum
of the contributions

\begin {equation}
u_1c+u_2c=1.710\times 10^{27}
\left(1-\sqrt{\frac{3}{r}}+\frac{0.0668
  r}{\sqrt{1+\frac{0.628}{f}}}\right)
  \frac{\dot m_d}{mr^3}\hspace {0.3cm} \rm{g/s^3}.
\end{equation}
The second contribution becomes more important than the first for $r$
larger than about 30. 
The contribution of hard radiation is proportional to the mass flow
rate in the corona and can be neglected in the soft state.  

Bremsstrahlung cooling is less important than Compton cooling when
Compton cooling is significant for a soft state disk underneath the corona. In
addition, 
conduction cooling is much less efficient than Compton cooling for the low
mass flow rates in the corona, as derived in the following.

\subsection{A limit to the mass flow rate in the corona}

To derive the mass flow rate in the corona, we evaluate the electron 
temperature for the case that Compton cooling $q_{\rm {cmp}}$ balances collisional
heating $q_{\rm{ie}}$. We use Eq.\ref{ADAF} to determine the relation
between $T_e$, $\dot m_c$, and $\dot m_d$ for a given distance $r$ and
$\alpha$. Here $f$=0.1 is again an appropriate approximation for which
the temperature difference $T_i$-$T_e$ in the formula for
$q_{ie}$ can be approximated by $T_i$ since the value of $T_e$ is
small compared to that for $T_i$, yielding
 
 \begin{eqnarray}\label{T_e}
 T_{\rm e}&=&4.010\times 10^{8}\alpha^{-2/5} \dot m_c^{2/5} \dot m_d^{-2/5}
   r^{1/5} \\
   &&\left(1-\sqrt{3\over r}
+ 0.025r\right)^{-2/5} \rm K. \nonumber 
\end{eqnarray}

Now, if Compton cooling brings the electron temperature
(Eq.\ref{T_e}), down below the
coupling temperature (Eq.\ref{T_cpl}) no two-temperature coronal flow
can exist for lower electron temperatures. This yields an upper limit
to the mass flow rate in the ADAF-type corona. For a given value $\dot
m_d$, this rate limit $ \dot m_c$ can be determined using 
$T_{\rm e}=T_{\rm cpl}$ ($f$=0.1). For a rate higher than this limit
the electron temperature would be higher, but the coupling temperature
would increase even more, preventing any coronal flow from existing.
The rate limit is given by

\begin{eqnarray}\label{eq-mdot_c}
 \dot m_{\rm c}&=&3.844\times 10^{-4} \alpha^{7/2} \dot m_d^{-3/2} r^{3/4}\\
 && \left(1-\sqrt{3\over r}
+ 0.025r\right)^{-3/2}. \nonumber
\end{eqnarray}
In Fig.\ref{f:mdot_c}, we show the resulting coronal mass flow rates and their 
dependence on both $\dot m_d$ and radius. 
For a mass flow rate in the disk $\dot m_{\rm d} = 0.1-0.2\dot
M_{\rm{Edd}}$, which is typical of a soft spectral state, the
accretion rate possible in the  ADAF-type corona is very low, below $10^{-3} \dot M_{\rm{Edd}}$. Only for a  
weak mass flow in the disk of around a few percent of $\dot
M_{\rm{Edd}}$, typical of an intermediate state, is the
maximal $\dot m_c$ slightly higher, at around 0.02$\dot
M_{\rm{Edd}}$.

We note that our
evaluation is very simplified and cannot give detailed
numbers, but it becomes clear that Compton cooling leads to a severe
reduction in the coronal mass flow. Narayan et al. (1998) indeed pointed already
out that bremsstrahlung cooling can reduce the coronal flow.

\subsection{Ion and electron temperatures affected by Compton cooling}

  \begin{figure*}
   \centering
   \includegraphics[width=6cm,height=5cm]{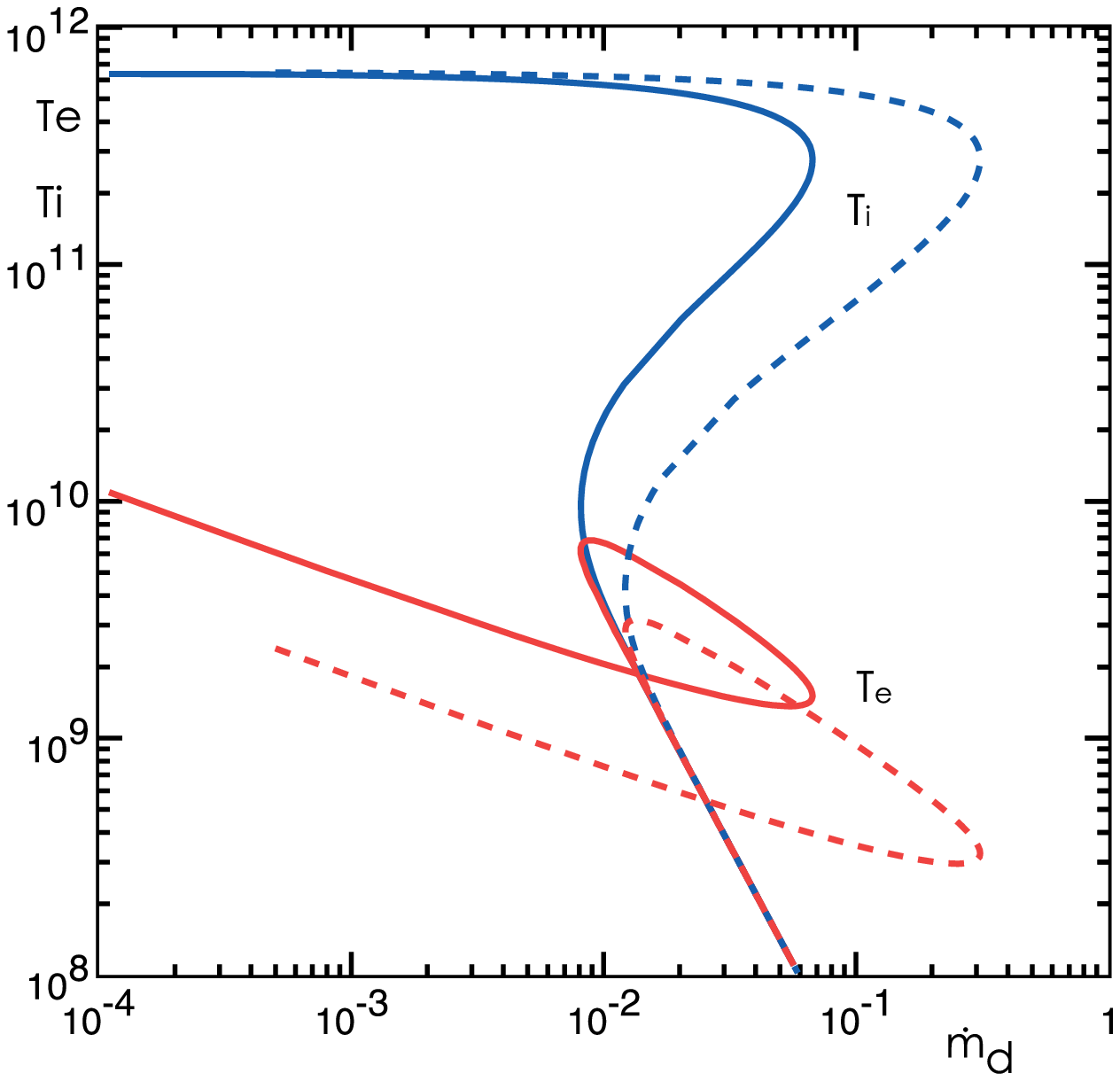}
\hspace{1.5cm}
   \includegraphics[width=6cm,height=5cm]{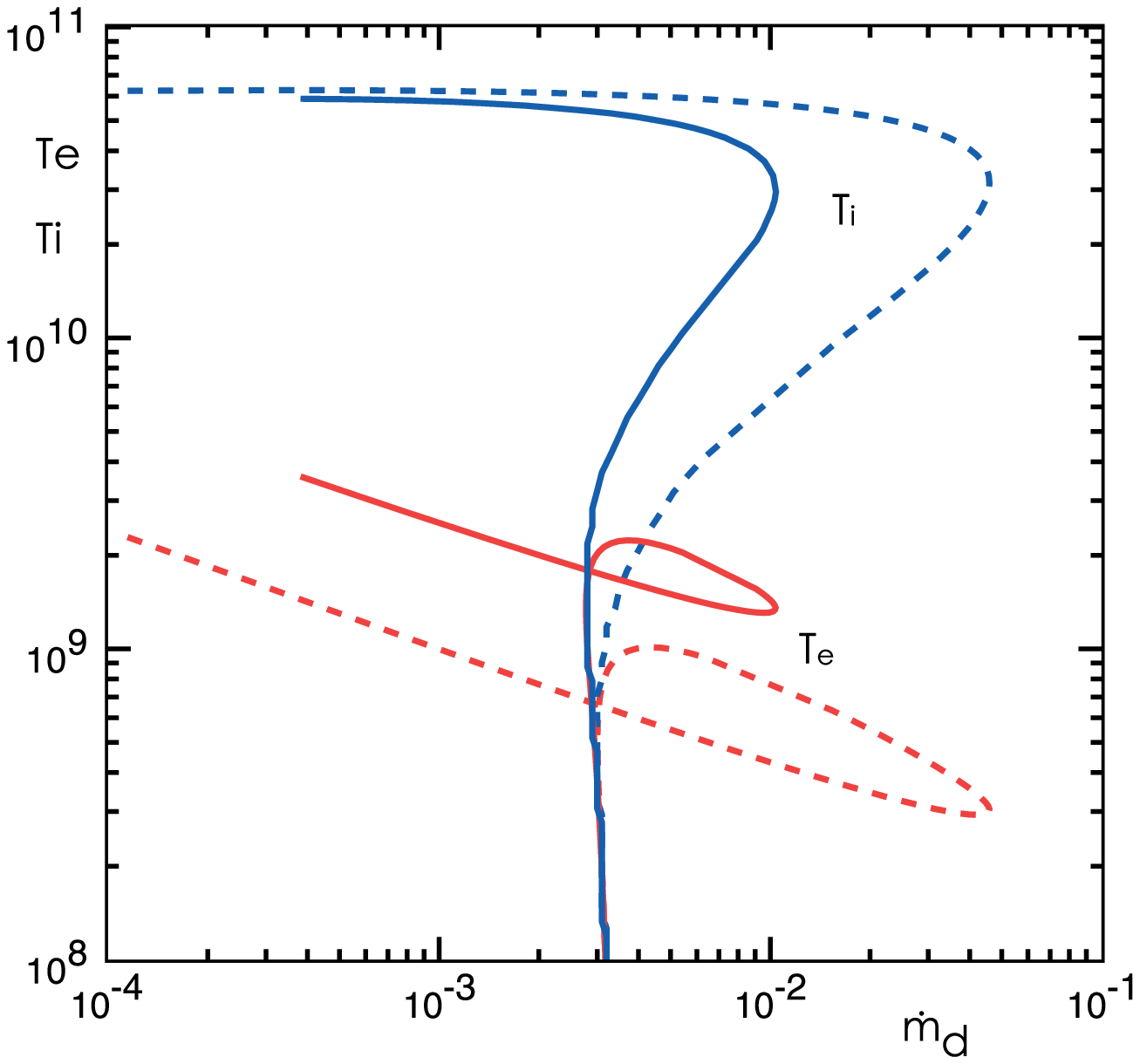}
      \caption   
{Ion and electron temperature $T_i$ and $T_e$ under Compton 
  cooling by photons from a disk underneath as function of disk mass flow rate
  $\dot m_d$. Solid lines: coronal mass flow rate $\dot m_{\rm{c}}$=0.01, 
  dashed lines:  $\dot m_{\rm{c}}$=0.001; left panel: $R=3R_{\rm{S}}$,
  right panel: $R=30R_{\rm{S}}$; $\alpha$=0.2}
   \label{f:TiTemdot_d}
   \end{figure*}

We determine how the mass flow in the disk affects the coronal mass
flow changing its nature from a typical two-temperature ADAF with $f$ close to
1 to a one-temperature flow with $f\rightarrow 0$. As stated 
above, we assume again that $(1-f)q^+=q_{ie}=q_{cmp}$,
that is all the energy obtained by collisions with ions ($q_{ie}$) is cooled 
by the Compton scattering of the disk photons. The disk mass flow is then

\begin{eqnarray}
\dot m_d&=&0.492(1-f)^{5/3}\alpha ^{\frac{7}{3}}c_3^{\frac{11}{3}}\dot m_c^{-\frac{2}{3}}{\epsilon'} ^{\frac{5}{3}}r^{\frac{1}{2}}\\
&&\left( 1-\sqrt{\frac{3}{r}}+\frac{0.0668 r}{\sqrt{1+\frac{0.628}{f}}} \right)^{-1}.\nonumber
\end{eqnarray}
This analytical formula is only valid for $T_e \ll T_i$. 
However, by evaluating the implicit formula without this approximation
one can study how ion and electron temperatures
change in depending on the cooling provided by a mass flow rate 
in the disk underneath. The result is shown in
Fig.\ref{f:TiTemdot_d}. With rising $\dot m_d$, the ion temperature
decreases monotonously, whereas the
electron temperature first decreases, then rises again, and finally
decreases again. The non-monotonic behavior of the electron
temperature is more pronounced for lower mass flow rates
in the corona and smaller radii. (If $r=3$ were approached, the
description of both the ADAF and Compton cooling would have to be modified.) 

For a given value $\dot m_c$, there is a critical rate $\dot m_d$
beyond which no solution exists. The critical rate $\dot m_d$ is
higher for lower $\dot m_c$, i.e. a higher mass flow in the disk
allows only a smaller mass flow in the corona.
Fig.\ref{f:TiTemdot_d} shows that there can be two solutions for
the same $\dot m_d$, a standard
ADAF solution with electrons poorly coupled to the ions, a high ion
temperature, and a  SLE-type solution with lower ion temperature (and
close to the black hole, a third one-temperature solution). The SLE
branch approaches a one-temperature solution. For low temperatures,
the densities in the coronal flow are higher and Compton cooling from
a certain amount of photons becomes more efficient, that is a very small
increase in $\dot m_d$ leads to a temperature decrease. For very low
ion temperatures the ADAF then becomes optically thick, e.g. for 
$T_i\le 5\times10^9$ and $\dot m_c=0.01$ and for $T_i\le
5\times10^8$ and $\dot m_c=0.001$ at $r=30R_S$. The effect of
cooling by the disk photons then differs.

We performed a stability analysis of the Compton cooling process,
following the procedure performed by Narayan \& Yi (1995b) who
considered other
cooling processes, other than cooling by a
disk underneath. We also find that the SLE branch is unstable, as
illustrated in 
Fig.\ref{f:TiTemdot_d} by the again increasing electron
temperature, following an initial decrease.

\subsection{Comparison with observations} 
The comparison with observations highlights a discrepancy.
Observations of X-ray binaries and Seyfert galaxies (in
soft spectral state) have found evidence of a coronal flow that is clearly
higher than the upper 
limits derived in our analysis. In their definition of spectral states,
Remillard \& McClintock (2006) indeed refer to an upper limit to the
non-thermal component of 25\% of the flux at 2-20 keV). The existence
of such a high power-law flux points to some additional heating of
electrons that is able to compensate the Compton cooling. 

In 2001, Merloni \& Fabian pointed out that the 
coronal energy content derived from observationally determined
electron temperatures and optical depth is insufficient to account for
the observed X-ray luminosities, and conclude that, if ion-electron coupling is
poor as assumed, an additional electron-heating mechanism, e.g. magnetic
flaring, is required.

We here show that coronal flows can be sustained even if up to 90$\%$
of the ion heating is drained to the electrons, but that owing to the
intricate dependence of the ion-electron coupling on both electron
temperature and density, a coronal flow can thus be sustained only for
a very low  mass flow rate, depending on the Compton cooling providing
a mass flow rate in the disk, that insufficient to account for the observed 
relatively high and hard X-ray radiation.

For such a process magnetic flares have been often proposed. This
explanation is supported by the observation of a 
non-thermal tail for several X-ray binaries in a soft state. For AGN,
only a few observations include this high-energy range in the
spectrum. For some sources, a cut-off seems to be indicated,
but the general situation is unclear.

\section{Observed non-thermal Comptonization}

Grove et al. (1998) used observations with OSSE on the
{\it{Compton Gamma Ray Observatory (CGRO)}} of seven transient galactic black
hole candidates to investigate the high-energy extension of the hard state. They
found three sources with hard spectra below 100 keV and exponential
cut-off consistent with thermal Comptonization, and four sources, with
relatively soft spectra, spectral indices
$\Gamma\sim$ 2.5-3, and no evidence of a spectral break. The
authors argued that photons from the disk in the high soft
state cool the electrons in an inner advection-dominated
Comptonization region. Our analysis of the Compton cooling effect is
consistent with this picture.

Observations of a hard power-law tail and several investigations of
spectra of Cyg X-1 have proven the existence of non-thermal 
Comptonization
(Gilfanov et al. 1999, Frontera et al. 2001, McConnell et al. 2002). 
McConnell et al. (2002) used observations from {\it{CGRO}} and
{\it{BeppoSAX}}  which extending over the energy range from 20 keV to
10MeV and derived constraints on the magnetic field in the hot flow 
based on the presence of a non-thermal tail. They
emphasized the similarity between black hole binaries in the hard
state and Seyfert galaxies and mentioned that similar tails could also 
be present in the AGN spectra. A detailed analysis of the broad-band
spectra of Cyg X-1 was presented by Ibragimov et al. (2005, see also
references therein). Beloborodov (1999) proposed a
situation different from the soft state and modeled the hard spectrum
of Cyg X-1 
suggesting, that during the hard spectral state a large fraction of
luminosity is released in a magnetic corona atop a cold accretion
disk, via compact bright flares, the magnetic energy is 
generated by the magneto-rotational instability. Poutanen \& Fabian
(1999) also argued that magnetic flares cause the
short-time spectral variability of Cygnus X-1.

For GRS 1915+105, evidence of non-thermal Comptonization was also 
found and discussed first by Grove et al. (1998), then Zdziarski et
al. (2001,2005). For GRO J1655-40, a high-energy cut-off during
the rising phase of an outburst and its disappearance in the high/soft
state was found by Joinet et al. (2008)
 ({\it{INTEGRAL}} observations along with {\it{RXTE}} and{\it{Swift}}
data). The studies of GX 339-4 by Motta et al. (2009) showed the
evolution of the high-energy cut-off during the brightening of the source
and also the final disappearance in the soft state.

Zdziarski et al. (2000) summarized the spectra of Seyfert
galaxies observed by the OSSE detector aboard {\it{CGRO}} and pointed
out that constraints on the form of the individual soft $\gamma$-ray
properties are rather weak owing to the limited quality of photon 
statistics. In the
case of the bright radio-quiet Seyfert galaxy, NGC 4151, the
spectrum above 50 keV is well described by thermal Comptonization.
The spectrum of the bright Seyfert galaxy MCG-6-30-15 was 
modeled 
with a relativistic reflection model in the range 0.5-200 keV by Chiang
\& Fabian (2011), but for the soft $\gamma$-ray range no information
is available. 

\section {Discussion}

Our results evoke a number of questions. Observations of
X-ray binaries may allow us to gain information on particular features of
the accretion flow geometry. For AGN, it is more difficult to determine
the spectral state of a source, since there are large 
uncertainties in e.g.
the bolometric luminosity measurements, as described 
by Raimundo, Fabian, Vasudevan et al. (2012).

\subsection{Dichotomy in the relation between ion and electron
  temperature}
The run of the  $T_i$ and $T_e$ curves in Fig.\ref{f:TiTemdot_d}
shows that for small distances two
temperature solutions are possible for the same value of Compton
cooling by photons originating from the disk underneath. This
indicates that different
combinations of $T_i$ and $T_e$ are possible for either an increasing
or a decreasing mass flow rate in the outer disk. During the
transition from hard to soft spectral state and the reverse
transition, the mass flow rates can be expected to allow such a 
temperature dichotomy in the hot flow. Stiele et
al. (2011) found in their study of GX 339-4 interesting differences in
the spectra when the source moves through both softening and hardening 
in the hardness-intensity diagram (HID), and argue that type B
quasi-periodic oscillations (QPO) are related to the electron
temperature distribution.

\subsection{Limitation of the hot coronal flow ?}
Dunn et al. (2010) show in their diagrams of disk fraction and power-law
fraction luminosities  how for several black hole binaries these
contributions change during the outburst cycles. In the soft state, the
power-law flux strongly fluctuates. The average amount of power-law flux
is in the range of a few up to ten percent of the total flux.
An even higher power-law flux of up to 17\% of the total flux was found 
in Swift/BAT observations of GX 339-4 during the 2010
outburst (Cadolle Bel et al. 2011, Shidatsu et al. 2011). This
flux decreased within a short time after outburst maximum. However,
the power-law flux observed for XTE J1752-223 during the 2009-2010
outburst (Swift/BAT observations), which is also a comparable fraction of
the total flux, did persist for the entire outburst duration (Nakahira et
al. 2012).

Within the picture developed here, a power-law flux higher than the
limitation by Compton cooling, which is around 1\% of the total flux,
should indicate that electrons are accelerated in another way, e.g. 
by magnetic flares. The magnetic flares otherwise
should yield non-thermal Comptonization, leading to a hard power-law 
tail in the spectrum up to energies higher than 200 keV. One might 
therefore expect that in the observations the two features appear 
together, i.e. that (1) the power-law
flux is  higher than the limitation found in our analysis and (2) there
is a hard power-law tail in the spectrum due to the presence of magnetic flares.

\subsection{Corona and disk during the intermediate state}
The two accretion flow configurations with 
either a disk reaching inward to the ISCO or an ADAF in the inner
region, and either dominant soft or hard radiation, are commonly
accepted. However, during the change from one of these states to the
other the accretion flow geometry is difficult to constrain. 

During outburst rise, and a rising mass flow from further
out, one can imagine that the inner edge of the truncated disk 
is continuously shifted inward, and that the soft radiation becomes 
dominant. During an outburst decline, the geometry of the accretion 
flow might be more 
complicated and depend on the process initiating the change from 
disk accretion to an ADAF. The efficiency of this
process, which is theoretically understood as a siphon process leading to
the evaporation of the disk (Meyer et al. 2000, Liu et al. 2002), 
has a maximum at a certain distance of probably a few hundred
$R_s$ (depending on parameters such as the ratio of gas
pressure to total pressure (Qiao \& Liu 2012)). This means that the
standard disk will first disappear there and that a gap might open 
between the outer standard disk and an inner weak disk. 
(In spectral fits, it might
be difficult to distinguish between the inner edge of
the standard disk and that of a weak inner disk.) The amount of mass in the
inner disk is continuously reduced by diffusion, but disappears only
slowly possible owing
to the re-condensation of gas from the corona into the inner disk
(Liu et al. 2006). The re-condensation only works as long as 
$\dot m \ge 10^{-3}$. So the weak inner disks are a natural feature of
the intermediate state, at the soft-to-hard state
transition. Observational evidence of a thermal component in the
spectra of quite a number of sources support this picture. The iron emission
lines then are the reaction of the inner accretion disk to irradiation.

For the intermediate state of Cyg X-1, Ibragimov et al. (2005)
discussed an Fe $K\alpha$ emission line. Observations with
{\it{Chandra, XMM-Newton}}, and {\it{Suzaku}} revealed relativistic 
iron emission lines for several X-ray binaries in the hard (probably  
hard-intermediate) state, and also for Seyfert AGN (Nandra et
al. 2007, Miller 2007). Additional
information is provided by {\it{Swift}} observations (Reynolds \&
Miller 2011).

The more complex accretion flow geometry in the intermediate
state does not contradict the fundamental picture of the two main flow
patterns, either a disk or an ADAF in the inner region.

\subsection{ Magnetic flares in the accretion disk?} 

If magnetic flares are important for the power-law flux in the soft 
state, the strength of the magnetic fields is of interest, and maybe 
also changes
in the magnetic field during the outburst cycles. Advective
concentration in the inner parts of 
accretion disks was investigated for binary stars by Meyer
(1996). The global, three-dimensional magnetohydrodynamic simulations
of Tchekhovskoy et al. (2011) show that a large amount of
magnetic flux can be transported to the center. This could lead to
magnetic flares in the inner region, the acceleration of electrons,
and a resulting power-law flux.  The different amounts of power-law
flux recently
observed for  GX 339-4 and XTE J1752-223, as discussed above, could
well be due to different amounts of flux being accumulated at the center
during different outbursts. Dunn et al.(2011) demonstrate in their 
analysis of the power-law flux during outbursts of the sources 4U 1543-47, 
GRO J1655-40, GX 339-4, H1743-322, XTE J1550-564, and XTE J1859+226 that there is a remarkable difference in the
variability of the power-law flux in the soft and in the hard 
spectral state. This
variability in the soft state could be caused by flare activity, which
is also a sign of magnetic fields.

\subsection{An effect of mass outflow from the corona? }

Accretion disk winds have been found for many galactic X-ray binaries 
in soft spectral state (Ponti et al. 2012), and 
especially the microquasar GRS 1915+105 (Neilsen et al. 2011). 
How would such a wind loss
influence the results for ADAF/coronal flows?

Wind loss leads to an inward successive decrease in the local mass 
accretion
rate and also takes up energy from the corona. The reduction
in the accretion rate can be approximately followed by our semi-local 
analysis with a radially decreasing rate. 

The impact of a reduced ion heating can be estimated by noting 
that the ADAF-type coronal flow is sustained until the coupling to the 
electrons takes away about 90\% of the ion heating rate. Even if wind
cooling were to take away 50\% of the gravitational energy released to
the ions, the ADAF character would still remain until electron
coupling took away the rest. Owing to the very sensitive dependence of 
the electron-ion coupling in this temperature range, such wind loss would
probably have a minor effect on the local $T_e-T_i$ relation and
the consequential limits to the coronal mass flow.

\section {Conclusions}

When broad iron emission lines are observed from a region close to an
accreting black hole, obviously both a coronal hot flow and a disk
irradiated by X-rays from this hot flow are present. Such an accretion 
flow geometry  either exists in the soft spectral state, with a 
standard disk +
weak coronal flow, or in the hard-intermediate state with a weak inner 
disk + strong coronal flow. 

In the first case, a standard disk in the soft state + corona, 
the coronal mass flow rate $\dot m_c$ is expected to be low, as the
result of the Compton cooling for an ADAF-type corona, e.g. 
$\dot m_c\le0.001$ for $\dot m_d=0.1$. The observation of a higher
coronal flux points to an
additional acceleration of the electrons in the corona, as might
result from flares of disk magnetic fields. Magnetic flares might yield a
tail of non-thermal radiation in the spectrum, as observed for
X-ray binaries in soft state. From energy considerations, Merloni 
\& Fabian (2001) pointed out the need for additional electron heating,
proposing that this is generated by magnetic flares. Support for this
idea also comes from observations of black hole binaries, especially
GRO J1655-40. The growing evidence of magnetic processes could be
important for the understanding of jet production.

In the second case, of a weak inner disk in the hard-intermediate 
state + corona, a coronal flow stronger than that in the soft state is 
possible. The more
complex accretion flow geometry, that is expected presumably near the
soft/hard spectral transition, might be present for quite some time as 
a transient phenomenon, until the weak inner disk disappears with a
decreasing accretion rate during the outburst decline. Differences in 
the relation between
ion and electron temperatures of the coronal flow between the
hard/soft and the soft/hard transition might be the cause of changes
in the variability character.

For X-ray binaries with observed broad iron emission lines, it is
possible to determine the spectral state and analyze the accretion
flow geometry. This is much more difficult for AGN (Raimundo, Fabian, 
Vasudevan et al. 2012). However, the possible appearance of broad iron
emission lines in different configurations of accretion flow geometry
allows us to understand that these lines are very common in
Seyfert-1 AGN (Nandra et al. 2007, Miller 2007).

\begin{acknowledgements}
{\rm{We thank Werner Collmar for information about $\gamma$-ray
    observations. 
   B.F. Liu thanks the Alexander-von-Humboldt Foundation for
     support}} 
{\rm{ and acknowledges support by the National Natural Science
 Foundation of China (grants 11033007 and 11173029) and by the National Basic
Research Program of China-973 Program 2009CB824800. }} 
\end{acknowledgements}

{}

\end{document}